\documentclass[aps,pre,floats]{revtex4}
\usepackage{graphicx}
\usepackage{amssymb}
\begin{document}
\title{Analytic approach to the evolutionary effects of genetic exchange}

\author{Elisheva Cohen} 
\affiliation{Dept. of Physics, Bar-Ilan  University, Ramat-Gan, IL52900 Israel}
\author{David A. Kessler}
\affiliation{Dept. of Physics, Bar-Ilan University, Ramat-Gan, IL52900 Israel}
\author{Herbert Levine} 
\affiliation{Center for Theoretical Biological
  Physics, University of California, San Diego, 9500 Gilman Drive, La
  Jolla, CA 92093-0319} 
\date{\today}
\begin{abstract}
We present an approximate analytic study of our previously introduced model of evolution including the
effects of genetic exchange.  This model is motivated by the process of bacterial transformation.  We
solve for the velocity, the rate of increase of fitness, as a function of the fixed population size, $N$.
We find the velocity increases with $\ln N$,  eventually saturated at an $N$ which depends on the strength
of the recombination process.  The analytical treatment is seen to agree well with direct numerical simulations
of our model equations.
\end{abstract}
\maketitle 

\section{Introduction}
Recombination of genetic information is an important strategy employed by biological systems to foster 
evolutionary novelty and to mitigate the adverse effects of deleterious mutations~\cite{review}. It is therefore critical to understand how the recombination details interact with factors such as mutation and finite population size so as to determine the overall Darwinian dynamics. Theoretical concepts and models can be  directly tested by comparison with laboratory-scale experiments on microorganisms~\cite{experiment-review}, especially those that can switch from asexual to sexual reproduction as a function of controllable conditions~\cite{nature,nature2}.
 
There has of course been a great deal of work on both the effects of the exchange of genetic information and on the evolution of sex~\cite{muller,kondrashov,barton,charlesworth}. Yet, analytically tractable approaches which can address the role of finite population size (and the resultant linkage of different loci) within a full genomic model are still lacking~\cite{rouzine2}. For example, most work to date focuses on systems with just two loci, although this limit is not at all suitable for the majority of microorganism systems. Including recombination in the simple landscape models ~\cite{evol-prl,evol-papers,rouzine} that have proven useful for asexual evolution~\cite{lab-evol}  is, we feel, the first step in this necessary direction.

The purpose of this paper is to present a detailed analytical investigation into a previously introduced~\cite{kessler}  model of recombination motivated by the phenomenon of genetic exchange by competent bacteria. Under proper conditions, many species of bacteria 
 can import snippets of DNA from the surrounding medium;
presumably these are then homologously recombined so as to replace the
corresponding segments in the genome~\cite{competence}.  Most
biologists are convinced that this process serves to enhance genetic diversity and thereby allows for better response to environmental  challenges faced by the colony~\cite{levin,redfield}. Our work addresses the conditions under which this type of genetic exchange is likely to be beneficial.

\section{Evolutionary Model}        
Our model~\cite{kessler}  consists of a population of $N$ 
individuals each of which has a
 genome of $L$ binary genes.  An individual's fitness $x$ depends additively on 
the genome $
x \ = \ \sum_{i=1}^L S_i$ with $S=0,1$.
Evolution is implemented as a continuous time Markov process in
which individuals give birth at the rate $x$ and die at random so as to
maintain the fixed population size.  Every birth allows for the
daughter individual to mutate each of its alleles with probability
$\mu _0$ giving an overall genomic probability of $\mu = \mu _0 L$. In addition, we add a process which mimics the aforementioned method of
recombination.  At the rate $f_s L$, an individual has one of its genes
deleted and instead substitutes in a new allele from the surrounding
medium; the probability of getting a specific $S$ is just its
proportional representation in the population.  This last assumption should be valid as long as the distribution of recently deceased
(and lysed) cells is close to that of the current population; this
should be the case whenever the random killing due to a finite carrying capacity is the most common reason for death.

The aforementioned Markov process is much too complicated to be solved exactly. In our previous work, we carried out a set of simulations to address the effect of finite values of $f_s$. We showed that 
at very small population sizes,
recombination has little effect, since there is no population
diversity upon which to act. At large $N$, the rate of evolutionary advance is much higher and again 
roughly independent of the recombination rate; this is because the effects of linkage disappear. This would occur even without recombination, albeit at values of $N$ that are unattainably large, even in viral experiments~\cite{evol-papers}. Most importantly, the population scale for the rise
is a strongly decreasing function of $f_s$, and hence
recombination at intermediate $N$ can give a dramatic speedup of the
evolution.  This basic result is
qualitatively consistent with recent experiments~\cite{nature,nature2} in
microorganism evolution which demonstrate an increase in the efficacy
of recombination as the population size is increased (starting from
small).

In order to approach this system analytically, we wish to derive an effective equation governing the fitness distribution of the population as a function of time. In the absence of recombination, it has been convincingly demonstrated that this can be accomplished by modifying the naive  mean-field theory (aka the Eigen-Schuster quasispecies equations~\cite{eigen-schuster}) by adding a cutoff on the birth rate if the population density near the leading edge goes below $P_c \equiv k/N$, for some $k$ of $O(1)$. The key to this idea, introduced independently in several different contexts~\cite{mf-dla,kepler,evol-prl,bd,klnature,prl}, is that the major effect of finite population size is to prevent the leading edge of most-fit individuals from spreading too far and too fast, as in reality there must be at least one individual (out of $N$) at a certain fitness for the equation to make sense. This notion leads to two different equations for the dynamics
of the fitness distribution function, depending on the size of $P$. 
First, if  $P$ is larger than $P_c$, we have
\begin{equation}
\frac{d P_x (t)}{dt}  =  (x   -\lambda) P_x (t) 
 +  \mu  \left[ \frac{(x+1)^2}{L} P_{x+1} (t) +  
(x-1) \left( 1-\frac{x-1}{L} \right)  P_{x-1}(t)  -xP_x (t) \right]   + {\cal S}_x
\label{MFE}
 \end{equation}
The first two terms reflect the birth-death process
and the genomic mutation respectively; the explicit form of the
mutation term arises from considering the probability of an individual
with fitness $x$ giving birth (rate $\sim x$), mutating ($\sim \mu$),
and hence going either up ($\sim (1-x/L)$, the number of currently
bad alleles) or down ($\sim x/L$, the number of good alleles). Alternatively, if $P$ falls below $P_c$ we drop the birth-death term $(x- \lambda ) P_x$ in the above equation. As already mentioned, it is only important to impose this cutoff on the {\em leading} edge of the population, not at the trailing edge.
Finally, $\lambda$ is a Lagrange multiplier arranged so as to maintain the total population size at $N$;
 $$
\lambda =  \frac{\int dx \, xP_x \theta(P_x-P_c) }{ \int dx \, P_x \theta(P_x -P_c)}
$$
For large $N$, $\lambda$ is essentially just the population mean fitness $\bar{x}$ and we will use this in what follows.

The last term, ${\cal S}_x$, reflects the recombination effect.  In our previous work, we verified that at all but smallest values of $N$ it was reasonable to assume that the subpopulation at some particular fitness $x$ had the same distribution of '0' and '1' alleles
at each locus. This assumption means that selecting at random an allele from the extracellular medium gives the site-independent chance $x/L$ of getting $S=1$ and $1-x/L$ of getting $S=0$. Following the above logic, this gives
 \begin{eqnarray}
{\cal S}_x &\equiv&  -  f_s L  \left[ \left( 1- \frac{\bar{x}}{L} \right) \frac{x}{L} P_x (t)
 + \frac{\bar{x}}{L} \left( 1-\frac{x}{L} \right) P_x (t) \right.
  -   \nonumber \\
  &\ & {}  \left.  \left( 1- \frac{\bar{x}}{L} \right) \frac{x+1}{L} P_{x+1} (t)
 + \frac{\bar{x}}{L} \left( 1-\frac{x-1}{L} \right) P_{x-1} (t) \right]  
\end{eqnarray}
The remainder of this paper is devoted to solving this equation, using the fact that $N$ is large. Of particular interest will be the typical scale of $f_s$ that is needed to recover the fast evolutionary advance expected in the no-linkage limit. 

\section{Approximate Pulse Solution}

We are interested in propagating solutions in which the fitness distribution function takes the form of a localized pulse with a mean fitness and a typical width~\cite{evol-prl,rouzine}. First, we will replace the spatial finite-difference equation by a PDE by taking the continuous space limit. Later, we will return briefly to the question of when this is quantitatively valid. For convenience, we will also rescale time by a factor of $L$, since as we will see the natural scale of velocity in the above equation is $O(L)$. This leads to 
\begin{eqnarray}
\dot{P}_x& = & P_x \left( \frac{(x-\bar{x})} {L} +
\frac{4x \mu }{L^2} -\frac{\mu}{L}+\frac{f_s}{L} \right) \nonumber \\
&+& P_x'   \left( \frac{2x^2\mu}{L^2}-\frac{\mu x}{L}
+f_s \left( \frac{x-\bar{x}}{L}
\right) \right) \nonumber\\
&+&\frac{P_x''}{2} \left( \frac{\mu x}{L}+\frac{f_s}{L}
\left( x+\bar{x}-2\frac{x\bar{x}}{L} \right) \right) .
\label{equation_cont}
\end{eqnarray}

In a system with translation invariance, a fixed velocity pulse would be an exact solution of the governing equation of the traveling wave form
$$
P_x (t) \equiv p(x-vt)
$$
Here, such a form will be approximately valid, to the extent that we can use a quasistatic approximation and treat the pulse shape as approximately constant in time. Later, we will discuss why this is valid at small velocity and how it breaks down at  very large $N$.  In detail, we set $x = \bar{x} (t) +z$, $v \equiv \frac{d\bar{x}}{dt}$ and obtain
\begin{eqnarray}
0 & = & p \left( \frac{z}{L} +
\frac{4(\bar{x}+z) \mu }{L^2} -\frac{\mu}{L}+\frac{f_s}{L} \right) \nonumber \\
&+& p'   \left(  v + \frac{2(\bar{x}^2 +2 \bar{x}z +z^2) \mu}{L^2}-\frac{\mu (\bar{x}+z)}{L}
+f_s  \frac{z}{L}  \right) \nonumber\\
&+&\frac{p''}{2} \left( \frac{\mu (\bar{x}+z)}{L}+\frac{f_s}{L}
\left( 2\bar{x} +z-2\frac{(\bar{x}+z)x}{L} \right) \right).
\label{pulse}
\end{eqnarray}
We will focus on the specific example of the speed at the midpoint of the genomic landscape, $\bar{x} =\frac{L}{2}$, treating it as a constant parameter; extensions to other values will be mentioned below. Ignoring the dependence of
the pulse shape on the time-dependent $\bar{x}$  is the technical manifestation of the aforementioned quasistatic approximation.  We can drop terms of order $\frac{\mu}{L}$ and $\frac{f_s}{L}$ as compared to $\frac{z}{L}$, since the width of the pulse will turn out to be much greater than unity for large $N$. Finally, we ignore the $z$ dependence in the diffusion constant as this purely mutational piece is smaller than the leading order (constant) contribution.  This then leads to our basic pulse equation
\begin{equation}
0=p_z \frac{z}{L}+p_z'\left( v+F\frac{z}{L}\right)
+\frac{F}{4}p_z''
\label{eqwkb}
\end{equation}
where $F=f_s+\mu$. 
The solution to this equation  must then be matched to the solution of the simpler equation that governs the region past the cutoff where $p < P_c$.

\section{WKB Solution - Leading Orders}
Our pulse equation is equivalent to a parabolic cylinder equation and so can be solved exactly and subsequently matched to the solution past the cutoff~\cite{elisheva}. This procedure would leave us with a complicated special-function  equation for the velocity, which would have to be solved either numerically or via asymptotic approximations valid at large $L$. We find it more transparent to derive the needed approximations directly from our equation. We first rescale our equation, 
defining $y-Fz/L$.  In terms of $y$ our equation reads
\begin{equation}
0=y p(y) +\frac{ v+y}{\Lambda} p'(y) +\frac{1}{4\Lambda^2}p''(y)
\label{eqwkby}
\end{equation}
where $\Lambda\equiv L/F^2$ and is assumed large.  Given that the highest derivative is multiplied by the
small factor $1/\Lambda^2$ it is natural to write down a WKB approximation for $p$, 
\begin{equation}
p(y)=Ce^{S/\Lambda}
\end{equation}
To leading-order
\begin{equation}
S' = 2\left(-v-y+ { \sqrt{ (v+y)^2-y }} \right)
\label{wkb_stag}
\end{equation}
and so with the convention that $S(0)=0$, we obtain
\begin{eqnarray}
S  &=&  \left[ -2vy - y^2 +
(v+y-\frac{1}{2})\sqrt{(v+y)^2-y} \right. \nonumber \\
&&-  v(v-\frac{1}{2}) - \left(v-\frac{1}{4}\right)\ln(1-4v) \nonumber \\ &\ & \left. +\left(v-\frac{1}{4}\right)
 \ln(-2v-2y+1-2\sqrt{(v+y)^2-y})
\right] \nonumber \\
\end{eqnarray}
$C$ is a normalization constant that needs to be determined by the integral condition, $\int_\infty^\infty p(z) dz =1$. The dominant contribution to the normalization integral arises from the region of the peak where we can use a quadratic approximation for $S$, $S\approx -y^2/2v$.  This gives us the very simple result
$C = \frac{1}{\sqrt{2 \pi Lv}}$. 

The velocity is determined by matching to the region to the right (R) of the cutoff point. The equation in this region is 
\begin{equation}
0=\left(v+y\right)p_R'(y) +\frac{1}{4\Lambda} p_R''(y)
\label{p<pcai}
\end{equation} 
the WKB solution to which is
\begin{equation}
p_R \simeq C_R e^{-\Lambda(4vy - 2y^2)}
\end{equation}
from which, using the condition $p_R(y_c)=P_c$ we obtain
\begin{equation}
p_R \approx e^{-\Lambda(4v(y-y_c) + 2(y^2-y_c^2))}
\end{equation}
Now, since $S'(y_c)$ for the pre-cutoff solution does not equal $S'(y_c)$ for the post-cutoff solution,
there is in general no way that the two WKB expressions can satisfy the matching condition on the first derivative.
The only resolution of this problem is if the pre-cutoff WKB solution breaks down before the cutoff, due to the presence of a turning point.
The turning point, $y_*$, is given by the discriminant condition 
$(v+y_*)^2-y_*=0$, so
\begin{eqnarray}
y_*&=&\frac{1-2v-\sqrt{1-4v}}{2}\ ,\nonumber\\
S'_*&\equiv& S'(y_*)= \sqrt{1-4v}-1  \ .
\end{eqnarray}
This allows us to calculate the amount of exponential decline from the peak at $S'=0$ to the turning point $S'_*$ by just plugging into the previous expression. 
\begin{eqnarray}
\ln{\frac{p(y_*)}{p(0)}}  &= &  \frac{L}{2F^2}\left[ 3v-1+\sqrt{1-4v} \right.
\nonumber \\ &&  \left. +(\frac{1}{4}-v)\ln(1-4v)\right]
\end{eqnarray}
Since, as we shall see, the turning point is in general close to $y_c$, this already allows us to calculate the leading order
(geometrical optics) expression for the velocity:
\begin{equation}
\ln P_c =  \frac{L}{2F^2}\left[ 3v-1+\sqrt{1-4v} 
  +\frac{1-4v}{4}\ln(1-4v)\right]
  \label{pc1}
\end{equation}
For small $v\ll 1$,  this equation reduces to
\begin{equation}
\ln P_c = -\frac{2L}{3F^2}v^3
\end{equation}
so that the velocity scales as $F^{2/3}(-\ln P_c)^{1/3}$, which is what we  obtained previously~\cite{evol-prl} in a purely-mutational model with very large $L$.  This is consistent with the observation that at $\bar{x}=L/2$, mutation and recombination act essentially
identically, the only difference being the addition $\mu y$ term in the diffusion operator.  For small $v$, $y_*$ is proportional to $v$, and so this extra
term does not contribute.  The new feature of the calculation not contained in the previous work is the presence of a limiting velocity of $1/4$.
We will discuss the significance of this particular value below.  It should be noted that the
above formula has the velocity achieving $1/4$ at a finite, though very small value of $P_c$, instead of the 0 value we would expect. Thus
according to this formula, the velocity is undefined for $P_c$'s smaller than this value. The futher
corrections we derive will rectify this anomaly.

As is usually the case with WKB, we need to treat the region close to the turning point $y=y_*$ more carefully if we wish to obtain an accurate formula for the solution. Since this is the region wherein one has to be match to the solution past the cutoff, this care is indeed necessary. We let the solution $p$ be given as 
\begin{equation}
p=e^{\Lambda\left(\sqrt{1-4v}-1)(y-y_*) \right) }\phi
\end{equation}
where we have taken out of $p$ the exponential dependence at the turning point, $S_*'$. Substituting this expression into the equation for $p$, it is easy to show that $\phi$ obeys the equation
\begin{equation}
\frac{1}{4\Lambda^2}\phi''(y) + \frac{y-y_*}{\Lambda}\phi'(y) + \sqrt{1-4v}(y-y_*)\phi=0
\label{airyplus}
\end{equation}
The normal turning point type solution ignores the first derivative term thereby resulting in an Airy equation for $\phi$. The solution then is
\begin{equation}
\phi(y)=C_1 \textrm{Ai}\left(\frac{y_*-y}{\delta}\right)
\end{equation}
where the length scale $\delta$ is given by
\begin{equation}
\delta\equiv \left(2\Lambda\right)^{-2/3}(1-4v)^{-1/6}
\end{equation}
With this, it is clear that the first derivative term is irrelevant as long as $\sqrt{1-4v}\delta$ is much larger than $O(1/\Lambda)$,
or equivalently $1-4v \ll 1/\Lambda=F^2/L$. Since $L$ is always taken to be large, we see that the Airy approach will work for a range of $N$ that becomes astronomically large. Beyond this point, one must resort to a much more complicated solution (involving parabolic cylinder functions; this pedantic exercise will not be presented here). We do mention in passing that it is possible to prove that  $v$ will not reach $1/4$ until $N$ actually reaches infinity.

Since $\delta \sim O\left(\Lambda^{-2/3}\right)$, it is clear that the logarithmic derivative of $p$ is still dominated by $\Lambda S'_*$.  Thus the only way
to achieve a match of the first derivatives is if to leading order $p$ vanishes.  Thus, $(y_*-y_c)/\delta$ must be close to $\xi_0=-2.338$, the location
of the first zero of the Airy function.  Thus gives an additional fall of $e^{-\Lambda S'_* \xi_0 \delta}$ to the magnitude of $p$ over the distance between
$y_*$ and $y_c$. This is the next order, $O(\Lambda^{1/3})$ contribution to the velocity relation, which now reads:
\begin{eqnarray}
\ln P_c &=&  \frac{L}{2F^2}\left[ 3v-1+\sqrt{1-4v} 
  +\frac{1-4v}{4}\ln(1-4v)\right] \nonumber \\
  &\ & {} + \left(\frac{L}{4F^2}\right)^{1/3}\left(1-\sqrt{1-4v}\right)\xi_0(1-4v)^{-1/6}
  \label{pc2}
\end{eqnarray} 

\section{Full WKB Solution}

So far, we have calculated the first two orders in the velocity for large $L$. A full WKB approach should, however, consider all the terms that do not vanish in the $L\to \infty$ limit.  To achieve such a solution,
we have to obtain the physical optics WKB expression, and match to the post-cutoff solution.  A full discussion of this
procedure applied to a related model is given in Ref. \onlinecite{chem2}.  Here we will be brief, outlining the steps
without additional commentary.  The physical optics WKB
solution is
\begin{equation}
p=C\left[\frac{(y+v)^2-y}{v^2}\right]^{-1/4}\left[\frac{1-2y-2v-2\sqrt{(y+v)^2-y}}{1-4v}\right]^{1/2}e^{\Lambda S}
\end{equation}
This expression is obtained in the standard manner by just going to next order in the WKB expansion. This then has to be matched to a more accurate solution near the critical point. Taking into account the modification due to the small first derivative term
in Eq. (\ref{airyplus}), we can derive 
\begin{equation}
p=C_1 \left[1 - 2\Lambda (y-y_*)^2\right] \textrm{Ai}\left(-\frac{y-y_*}{\delta} + 2\Lambda\delta^2\right)
\end{equation}
The matching yields
\begin{equation}
C_1 = C \frac{2\sqrt{\pi v}\delta^{1/4}}{(1-4v)^{3/8}} e^{\Lambda S_*}
\end{equation}
We now have to match this to the post-cutoff solution, which has a logarithmic derivative (w.r.t. $y$) of $-4\Lambda (y_c + v)$ to leading order.
As noted above, this fixes the location of $y_c$:
\begin{equation}
y_c = y_* -\xi_0 \delta + 2\Lambda \delta^3 - \frac{1}{\Lambda (1-\sqrt{1-4v}) }= y_* -\xi_0\delta + \frac{1}{2\Lambda\sqrt{1-4v}} - \frac{1}{\Lambda (1-\sqrt{1-4v})}
\end{equation}
where the third term comes from the shift in the argument of the Airy function, and the last term from the small
distance from the zero of the Airy.  The condition that $p(y_c)=P_c$ then gives
\begin{equation}
P_c = C_1 \textrm{Ai}'(\xi_0) \frac{1}{\Lambda (1-\sqrt{1-4v})} e^{S'_*\left(-\xi_0\delta + \frac{1}{2\Lambda\sqrt{1-4v}} - \frac{1}{\Lambda(1-\sqrt{1-4v})}\right)}
\end{equation}
Putting this all together gives us our final expression for the velocity:
\begin{eqnarray}
\ln P_c &=&  \frac{L}{2F^2}\left[ 3v-1+\sqrt{1-4v} 
  +\frac{1-4v}{4}\ln(1-4v)\right] \nonumber\\
  &\ & {} + \left(\frac{L}{4F^2}\right)^{1/3}\frac{\left(1-\sqrt{1-4v}\right)\xi_0 }{(1-4v)^{1/6} }\nonumber \\
  &\ & {} + \ln \left( \left[\frac{16F}{L^2}\right]^{1/3} \frac{\textrm{Ai}'(\xi_0)}
  { (1-4v)^{1/6}(1-\sqrt{1-4v})}\right) + 1 - \frac{1-\sqrt{1-4v}}{2\sqrt{1-4v}}
 \label{pc3}
\end{eqnarray} 
 We compare our current expression for the velocity with numerical simulation data obtained by solving the original time-dependent pde and measuring the pulse speed when the mean fitness passes $L/2$. The results are presented in Fig. 1, where we in addition display the lower order approximations, 
 Eqs. (\ref{pc1}) and (\ref{pc2}).  We see that these additional contributions are indeed significant, even for
 the relatively large $L$ employed.
 
 \begin{figure}
\includegraphics[width=0.4\textwidth]{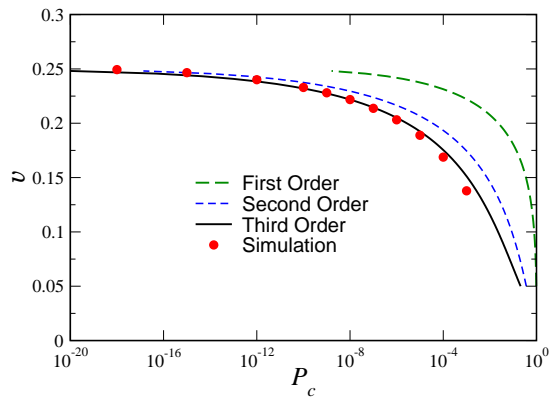}
\caption{Comparison of the analytic approximation for the velocity at $\bar{x}=L/2$ as a function of the cutoff $P_c$, Eq. (\protect{\ref{pc3}}), ("Third Order") with
the results of direct simulations of the time-dependent spatially-discrete equation, Eq. (\protect{\ref{MFE}}).  In addition, we show
the lower order approximations, Eqs. (\protect{\ref{pc1}}) ("First Order") and (\protect{\ref{pc2}}) ("Second Order"). The
parameters employed are: $L=1000$, $F=2.1$. (color online)}
 \end{figure}
 
\begin{figure}
\includegraphics[width=0.4\textwidth]{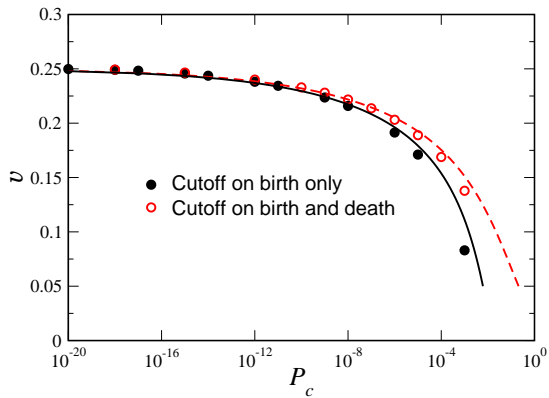}
\caption{Velocity vs. the cutoff, $P_c$, when only birth and not death is excluded beyond the cutoff.  For comparison, the
velocity from the original model with both birth and death excluded beyond the cutoff, is presented as well.  The results
of simulation (circles) as well as the analytic WKB prediction (lines) are displayed.  Parameters are $L=1000$, $F=2$.(color online)}
\label{birthfig}
\end{figure}

 One interesting fact that emerges from this analysis concerns the extent to which the selected velocity depends on the nature of the
  imposed cutoff. Recall that our cutoff dropped the entire birth-death term in the region past $P_c$. Since the first two terms in Eq.
   \ref{pc3} arose simply by requiring that $p$ approach {\em zero} at the cutoff, these would be unchanged had we employed a
    different treatment, say dropping only the birth term. This is also true of the $O(\ln L)$ piece.  Only the $O(1)$ contribution is
    modified.   In detail, in we were to drop only the birth term, the distance of $y_c$ to the zero of the Airy function is modified,
    and is now $-\frac{1}{\Lambda \sqrt{4-2\sqrt{1-4v}-4v}}$ so that the velocity-cutoff relation now reads:
   \begin{eqnarray}
\ln P_c &=&  \frac{L}{2F^2}\left[ 3v-1+\sqrt{1-4v} 
  +\frac{1-4v}{4}\ln(1-4v)\right] \nonumber\\
  &\ & {} + \left(\frac{L}{4F^2}\right)^{1/3}\frac{\left(1-\sqrt{1-4v}\right)\xi_0 }{(1-4v)^{1/6} }\nonumber \\
  &\ & {} + \ln \left( \left[\frac{16F}{L^2}\right]^{1/3} \frac{\textrm{Ai}'(\xi_0)}
  { (1-4v)^{1/6}(\sqrt{4-2\sqrt{1-4v}-4v})}\right) + \frac{1-\sqrt{1-4v}}{\sqrt{4-2\sqrt{1-4v}-4v}} - \frac{1-\sqrt{1-4v}}{2\sqrt{1-4v}}
 \label{pc4}
\end{eqnarray} 
In Fig. \ref{birthfig}, we plot this modified velocity together with the original.  We see that the general features of the two
models are similar, and that the modification results, for a given $v$, in an effective decrease in $P_c$ by a factor of about 10.
This could be absorbed in our phenomenological parameter $k$ relating $P_c$ to the $N$ in the stochastic model. 

Let us return to the scaling implied by the analytic formula. From the above, it is clear that the large $N$ limit in which $v$ asymptotes to $1/4$ is defined by
\begin{equation}
N> N_{cr} \sim e^{\frac{L}{8F^2}}
\end{equation}
This is a very strong function of $F$; one can reach the large population limit velocity at much smaller size by making a relatively modest increase in $F$. In this model, this behavior is what is responsible to the efficacy of recombination in improving the rate of evolution. Essentially, fitness improvement rates will be limited by clonal interference as $N$ becomes moderately large; different beneficial mutations arise in different lineages and cannot be jointly selected for. This can only be overcome if the variance of the population is so large that all different combinations of multiple mutants are present simultaneously or if recombination  succeeds in collecting these different mutations in a common line. This reduces the needed width, which is proportional to $\ln{N}$,  by a inverse factor of $F^2$. 

Now, the fact that mutation and recombination act similarly (adding up to give $F$) is a feature of the fact that we have chosen to do our calculation at the symmetric point $\bar{x} = L/2$. If the mean fitness is above the midpoint, the effects of $\mu$ and $f_s$ diverge. In particular, if we return to the original pulse equation (4), we see that there is a drift term proportional to $\mu$ that increases with $\bar{x}$; this term is merely the fact that in this part of the landscape most mutations are detrimental. Crucially, there is no such term proportional to $f_s$ arising from the recombination process. Thus, one cannot in general get to large positive velocity by increasing the mutation rate; the bias will win out. One must resort to a recombination strategy; there will be no recombination load as long as we assume (as we have) that genes are entities that cannot be broken apart by recombination. 

Looking more closely at what happens away from the symmetric point $\bar{x}=L/2$, for simplicity we drop the mutation term and concentrate solely on the
recombination.  There is now an addition term $Fz(1-2\alpha)/L p''(z)=(F^4/L^2)p''(y)(y(1-2\alpha)/F)$ in the steady-state equation.  
The calculation is similar; note however that now the equation is no longer a parabolic cylinder equation and no exact solution is possible; nonetheless
our WKB treatment still works.  Qualitatively the picture is the same.  The additional term in the equation is irrelevant
to small velocities, becoming more important as the velocity increases. The most interesting question is what happens to the limiting velocity.  The limiting velocity
is given in general by the point where $dS'_*/dv$ diverges.  A straightforward, but messy, calculation yields the result
\begin{equation}
v_{\infty}=\alpha(1-\alpha)\left(\frac{F}{1-2\alpha}\right)\left(1-\sqrt{1-2(1-2\alpha)/F}\right)
\end{equation}
where $\alpha\equiv \bar{x}/L$.  Note first that this is the "binomial" answer, $\alpha(1-\alpha)$, times a correction factor which is a function
of the ratio $F/(1-2\alpha)$. For large $F$, this correction factor approaches unity, and we recover the "binomial" answer.  This is consistent
with our observation~\cite{prl} that the effect of the recombination is to drive the system to the binomial distribution.  For $\alpha>1/2$, the
limiting velocity is less than the binomial result, whereas for $\alpha<1/2$ it is always greater.  For $F<2(1-2\alpha)$, our formula does not
predict any limiting velocity.  This just means that the velocity is controlled by the right boundary at $y=1-\alpha$ and not by the cutoff. 
We have verified our formula numerically for a number of different $\alpha$.  

It is worthwhile considering the origin of this deviation from the "binomial" answer for infinite $N$, in light of the easily verified fact that the binomial distribution with velocity  $\alpha(1-\alpha)$
is an {\em exact} solution of the fully time-dependent equations in the absence of a cutoff, for any value of $F$. The answer is that we are seeing
an effect of the breakdown of the quasi-static approximation for large $v$'s, of order $1$ in our scaling.  Thus, when $\alpha<1/2$, when the dynamic
interface is accelerating, the velocity does not have a chance to reach the quasistatic answer before it has to move on, and hence the dynamic
velocity is less than the quasi-static prediction.  Similarly, for $\alpha>1/2$, the interface is decelerating, and hence the velocity is greater than
the quasistatic prediction.  Only when $v \ll 1$ is the quasistatic approximation {\em quantitatively} valid; this limitation also emerges if one just estimates the terms being dropped in the original pulse equation. We do not know at present of an analytical method which can go beyond this limitation,

A last issue that is worth mentioning is the use of the space continuum approximation for our evolution system originally defined on a lattice. We have shown elsewhere that lattice effects can become important if the effective diffusion constant (here equal to $F/4$) is small. Out interest here is in the effect of taking $F$ fairly large by having a significant recombination rate; in this range, the continuum approximation is quantitatively justified. This can of course be seen a posteriori by the agreement between the calculations and the numerical velocity data. 

\section {Discussion}
This paper has been devoted to an analytic treatment of a previously developed model for evolution at finite population size with both mutation and one specific type of genetic recombination. Our results indicate that the rate of evolution can be dramatically speeded up by recombination as we approach the answers that would pertain in the infinite population size limit, where all genomes are almost immediately populated and selection alone accounts for the (unnormalized) speed of $L/4$. If mutations were strictly unbiased, similar effects could be had by a greatly increased mutation rate. But, as soon as we move up the fitness landscape towards the eventual equilibrium state, deleterious mutations are more common and increasing the rate increases the mutational load. In our model, there is no recombinational load. 

In order to see the effects of recombination, our basic rate of recombination events per gene, denoted by $f_s$ must be order $\sqrt{L/\ln{N}}$. In comparing to real systems, $L$ is the number of genes contributing to possible fitness changes in a novel environment and $N$ is of course the population size.  The rate itself is measured with respect to the differential birth rate increase due to the fixation of one beneficial allele. Our feeling is that it may be premature to try to compare this theory quantitatively to any specific set of experiments, but nonetheless the basic trends should be verifiable in microorganism experiments. 

Future work must address a number of points that are absent in our simple model but presumably present in the real microbial world. First, is the role of lethal mutations in biasing the genetic composition of the environment; some deaths are due to bad genes and these could be picked up by DNA importation. This is a form of recombination load, i.e. a bias towards the negative in the recombination process. A second issue along the same lines concerns the fact that homologous recombination may occur in the middle of a coding sequence and will bring together incompatible fragments. In our model this would appear as an epistatic interactions among the sites. Finally, our analytic method assumes we can write down an equation  solely in terms of the phenotype distribution $P_x (t)$. This clearly breaks down at small $N$ where different alleles have very different population distributions and we do not as yet have a method which can reach into this region. 

\acknowledgments{EC and DAK acknowledge the support of the Israel Science Foundation.  The work of HL has been supported in part by the NSF-sponsored Center for Theoretical Biological Physics (grant numbers PHY-0216576 and PHY-0225630).


\begin{thebibliography}{99}

\bibitem{review} For a review, see S. P. Otto and T. Lenormand,
 Nature Reviews Genetics \textbf{3}, 252 (2002). See also W. P. C. Stemmer, Nature \textbf{370}, 389 (1994); E~. Baum, D.~ Boneh and C.~Garrett, Evol. Comp. \textbf {9}, 93 (2001).

\bibitem{experiment-review} S.F. Elena and R. E. Lenski,  Nat. Rev. Genet. \textbf{4} 457�69 (2003)

\bibitem{nature} N. Colegrave, Nature {\textbf 420}, 664 (2002).

\bibitem{nature2} M. R. Goddard,  et al. Nature \textbf{434} 636-640  (2005). 

\bibitem{muller} H. J. Muller,  Mut. Res. {\textbf 1}, 2 (1964);
 J. Felsenstein,  Genetics {\textbf 78}, 737 (1974).

\bibitem{kondrashov} A. S. Kondrashov,  J. Hered. \textbf {84}
, 372 (1993).


\bibitem{barton} N. H. Barton,  Genet. Res. \textbf {65}, 123 (1995);
 S. P. Otto and N. H. Barton,  Evol. \textbf {55}, 1921 (2001).

\bibitem{charlesworth} B. Charlesworth, Genet. Res.  {\textbf 63},
 213 (1994); W. R. Rice,  Nat. Rev. Genet. \textbf {3}, 241 (2002).

\bibitem{rouzine2} For a different approach, see I.~ M.~Rouzine and J.~M.~Coffin, Evolution of HIV under Selection and Weak Recombination, Genetics (in press).

 \bibitem{evol-prl}L.~S.~Tsimring, H.~Levine, and D.~A.~Kessler, \prl
\textbf{76}, 4440 (1996).

\bibitem{evol-papers}
 D.~A.~Kessler, D.~Ridgway, H.~Levine, and L.~Tsimring,  J.
Stat. Phys. \textbf{87}, 519 (1997).

\bibitem{rouzine}I.~M.~Rouzine, J.~Wakeley, and J.~M.~Coffin, 
Proc. Nat'l. Acad. Sci. \textbf{100}, 587 (2003).

\bibitem{lab-evol} I. S. Novella et al.,  Proc. Nat. Acad. Sci.
 \textbf {92}, 5841 (1995); R.  Miralles, A. Moya and S. F. Elena,  J.
 Virol. \textbf {74}, 3566 (2000);
 R. E. Lenski, M. R. Rose, S C. Simpson and S. C. Tadler,  Am. Nat.
  \textbf {138}, 1315 (1991).

\bibitem{kessler} E.~Cohen, D.~A.~Kessler and H.~Levine, 
\prl \textbf{94}, 098102 (2005).

\bibitem{competence} D. Dubnau, Ann. Rev. Microbiol. {\textbf 53},
 217 (1999).

\bibitem{levin} B. R. Levin and C. T. Bergstrom,  Proc. Nat'l. Acad. Sci.
 \textbf{97}, 6981 (2000).

\bibitem{redfield} For a somewhat more skeptical view, see R. Redfield, 
 Nature Rev. Genet. \textbf {2}, 634 (2001).

\bibitem{eigen-schuster}M.~Eigen, Naturwissenschaften \textbf{58}, 465 (1971), 
M.~Eigen, J. McCaskill and P. Schuster, Adv. Chem. Phys. \textbf{75}, 149 (1989).


\bibitem{mf-dla} E.~Brener, H.~Levine, and Y.~Tu, \prl \textbf{66}, 1978
(1991).
 
\bibitem{kepler}T.~B.~Kepler and A.~Perelson, Proc. Nat. Acad. Sci. \textbf{92}, 8219 (1995).

 \bibitem{bd}E. Brunet and B. Derrida, {\pre} \textbf{56}, 2597  (1997).
 
 \bibitem{klnature}D.~A.~Kessler and H.~Levine,  Nature \textbf{394}, 556 (1998).


\bibitem{prl} E.~Cohen, D.~A.~Kessler and H.~Levine, \prl \textbf{94} 158302 (2005).

\bibitem{elisheva} E. Cohen, unpublished.

\bibitem{chem2} E.~Cohen, D.~A.~Kessler and H.~Levine, ``Front Propagation Dynamics up a Reaction-rate Gradient", cond-mat/0508663.

\end{thebibliography}
\end{document}